# Photoionization and Photoelectric Loading of Barium Ion Traps

A.V. Steele, L. R. Churchill, P. F. Griffin and M. S. Chapman
*School of Physics, Georgia Institute of Technology, Atlanta, Georgia 30332-0430*
(Dated: March 22, 2007)

Simple and effective techniques for loading barium ions into linear Paul traps are demonstrated. Two-step photoionization of neutral barium is achieved using a weak intercombination line ($6s^2\ ^1S_0 \leftrightarrow 6s6p\ ^3P_1$, $\lambda$ = 791 nm) followed by excitation above the ionization threshold using a nitrogen gas laser ($\lambda$ = 337 nm). Isotopic selectivity is achieved by using a near Doppler-free geometry for excitation of the triplet $6s6p\ ^3P_1$ state. Additionally, we report a particularly simple and efficient trap loading technique that employs an inexpensive UV epoxy curing lamp to generate photoelectrons.

Ion traps are an important technology for precision metrology with applications to frequency and time standards [1] and tests of fundamental physics [2, 3]. Ion traps also represent a leading candidate for quantum information processors and quantum simulators [4].

Ion traps are typically loaded by ionizing neutral atoms in the trapping region using electron beam bombardment. Recently, several groups [5-12] have demonstrated optical-based techniques for creating the ions that employ photoionization (PI) of neutral atoms. These techniques offer much greater loading efficiencies compared with electron impact ionization and also minimize charge build-up on insulating surfaces from the electron beam. Charge build-up can destabilize the trap and is thought to contribute to micromotion heating [13]. Minimizing charge build-up is particularly important for loading smaller chip-based ion traps [14] as well as for ion cavity QED applications that require dielectric mirrors in close proximity to the ion trap [15, 16]. We are particularly motivated to reach the cavity QED strong coupling regime with trapped ions, which requires mirror separations of < 1 mm for strong optical transitions. Photoionization loading also offers the advantage of isotope selective loading [17] and simplifies the ion trap apparatus by eliminating the necessity of the electron beam.

In this report, we demonstrate photoionization loading of a barium ion trap. Previous work in photoionization ion trap loading with Ca, Mg, Sr, Yb and Cd [5-12] have all employed two-step excitation beginning with a strong dipole allowed S-P transition. In our case, the strong transition ($^1S_0 \rightarrow\ ^1P_1$, $\lambda$ = 553 nm) requires an expensive dye laser, and instead we employ a weak intercombination transition that can be excited with a simple extended cavity diode laser (ECDL). We show the isotopic selectivity of the photoionization loading technique and measure the loading rate under a variety of experimental conditions. Additionally, we demonstrate a particularly simple loading technique that employs a UV halogen lamp to generate photo-electrons which then ionize the neutral atoms. Although this latter technique relies on electron bombardment, and hence can lead to charge build-up, it obviates the need for an electron beam source.

The relevant atomic structures of neutral and singly ionized Ba can be seen in Fig. 1. From the ground state of Ba, the $6s6p\ ^3P_1$ state is excited with a tunable single-frequency diode laser operating at 791 nm. From the $^3P_1$ state, the atom is excited to the continuum with an $N_2$ laser operating at 337 nm. Preferential isotopic loading is achieved by tuning the 791 nm laser to be resonant with the desired isotope and utilizing a Doppler-free geometry. Using this technique, all the stable isotopes with abundances greater than 1% are loaded, including the even isotopes ($^{134}$Ba, $^{136}$Ba, $^{138}$Ba), and the odd-isotopes ($^{135}$Ba, $^{137}$Ba).

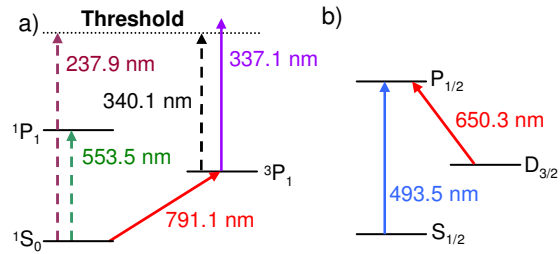

FIG. 1 (Color Online): Relevant energy levels of a) neutral Ba and b) singly ionized Ba. Solid lines indicate transitions used in this experiment. Dashed lines are included for reference.

A 4-rod linear Paul trap is employed for these investigations, similar to that described in [18]. The trap consists of four stainless steel rods of diameter 0.8 mm whose centers are arranged on the vertices of a square with a side of length 1.85 mm. A 6.6 MHz rf potential (50 $V_{rms}$) is applied to two of the diagonally opposed pins. The high voltage is generated using a copper helical resonator [19] (Q = 67) to step-up a pre-amplified signal from an rf oscillator. The remaining two rods are grounded. Axial confinement is provided by DC voltages (typically 7 V) applied to endcap electrodes consisting of stainless steel cylinders coaxial with the trap ground electrodes. The endcap electrodes are separated by 1 cm and are electrically insulated from the rods by thin polyimide tubes. This combination of trapping fields yield a radial secular frequency of $\nu_r$ = 626 kHz and an axial secular frequency of $\nu_a$ = 272 kHz. These frequencies are measured by observing heating of the trapped ions when the frequency of an external rf driving field is tuned to one of the secular frequencies.

A neutral barium atomic beam is created using a resistively heated tube oven formed from 0.025 mm thick spot-welded tantalum foil. The tube is 2 mm in diameter and 8 mm long with a 4 mm slit opening and requires a current of 3.8 A to reach the typical operating temperature of 300 ºC. The oven emits an effusive, uncollimated atomic beam that is directed at the ion trap situated ~1 cm above.

Laser cooling and imaging of the ions is accomplished using the transitions shown in Fig. 1(b) with laser light at 493 nm and 650 nm. The $S_{1/2} \rightarrow P_{1/2}$ transition is excited using a commercial doubled diode laser system (Toptica SHG-100) operating at 493 nm. The $P_{1/2}$ state can also decay into the metastable $D_{3/2}$ state ($\tau$ = 80 s) [20] with a 25% branching ratio [21], and a commercial ECDL (Toptica DL100) at 650 nm is used to repump the population out of this state. A weak magnetic field (1 G) is applied in order to prevent the formation of time-independent dark states in the $D_{3/2}$ Zeeman manifold [22].

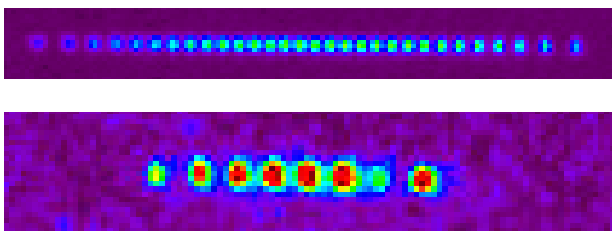

FIG. 2 (Color Online): Upper: False color image of chain of 30 $^{138}$Ba ions. Lower: Seven $^{138}$Ba ions and one $^{136}$Ba ion. Inter-ion spacing of the central ions is 12 μm in upper chain and 18 μm in lower chain.

The laser-cooled ions are observed by collecting the ion florescence using an NA = 0.28 objective; the collected light is either imaged onto a CCD camera or sent to an eyepiece for viewing. The imaging resolution is 3 μm, which is much smaller than the typical ion spacing (10-20 μm), making individual ions easily resolvable. Images of typical ion chains are shown in Fig. 2. For typical laser detunings (a few tens of MHz red detuned from the transition resonance), the detected signal corresponds to ~80 photons/ms/ion.

As briefly described above, loading the ion trap using photo-ionization of neutral Ba involves a two-step excitation. From the ground state of neutral Ba, the 6s6p $^3P_1$ state is excited with a tunable ECDL operating at 791 nm. An absolute frequency reference for this laser is obtained by saturated absorption spectroscopy in a heated barium vapor cell. Typical signals using standard frequency modulation techniques are shown in Fig. 3. Absorption lines and corresponding error signals are observed for the isotopes $^{135}$Ba through $^{138}$Ba. The 791 nm light is focused down to a waist of 100 μm at the trap center and is oriented perpendicular to the atomic beam to minimize Doppler shifts.

The second step of the photoionization entails excitation of the atom from the $^3P_1$ state to the continuum, requiring a wavelength <340.1 nm. A nitrogen gas laser operating at 337.1 nm provides a convenient and inexpensive solution, allowing excitation to 30 meV above the ionization threshold. A commercial laser (SRS NL 100) is employed that provides 170 μJ pulses with a width of 3.5 ns at a rate of 20 Hz. The laser beam is focused to a beam waist of 900 μm and counter-propagates with the 791 nm laser.

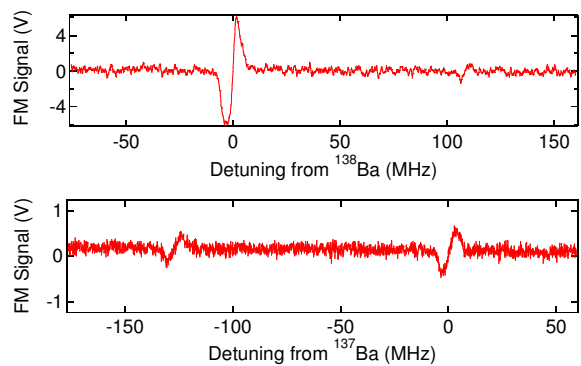

FIG 3 (Color Online): FM signals of the $^1S_0 \rightarrow ^3P_1$ transitions obtained from saturation spectroscopy in a Ba vapor cell. Upper: $^{138}$Ba and $^{136}$Ba signals. Lower: The F = 5/2 $^{135}$Ba and $^{137}$Ba signals. The zero of frequency here is detuned by +1.8 GHz from the $^{138}$Ba signal. The F = 3/2 and F = 1/2 $^{135}$Ba and $^{137}$Ba signals (not shown) are -0.8 GHz and -2.7 GHz detuned from the $^{138}$Ba line.

We first investigate the photoionization loading of the even isotopes, $^{134}$Ba, $^{136}$Ba, and $^{138}$Ba as a function of the detuning of the 791 nm laser beam, measured with respect to the $^{138}$Ba absorption line. The loading rates for these three isotopes are shown Fig. 4. The loading rates are determined by loading a small number of ions for a predetermined photoionization period. Following extinction of the photoionization lasers, the cooling lasers are turned on to cool the ions into a chain, allowing counting of the individual ions. This process is repeated several times, and the average loading rate is determined by dividing the total number of loaded ions by the total loading time. The uncertainties are determined assuming Poissonian statistics. The different isotopes are dis-criminated by comparing the ion image intensity at different detunings of the cooling lasers.

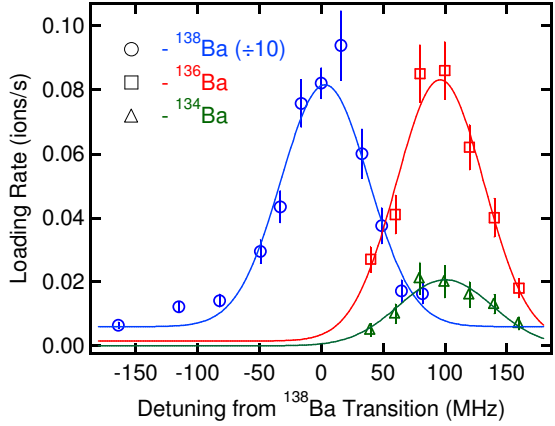

FIG 4 (Color Online). Loading rates of the even isotopes $^{138}$Ba, $^{136}$Ba and $^{134}$Ba vs. the detuning of the 791 nm laser. Data for isotope $^{138}$Ba are scaled (÷10) to fit on the graph. The solid lines are Gaussians fits with offsets corresponding to the background loading rate with the 791 nm light turned off.

The data in Fig. 4 are taken with a 791 nm laser power of 100 μW, which corresponds to a peak intensity, $I$ = 325 mW/cm$^2$ and a peak Rabi rate $\Omega_{Rabi}/2\pi = \Gamma/2\pi \sqrt{I/I_{sat}}$ = 7.6 MHz, where $I_{sat}$ = 0.014 mW/cm$^2$ is the saturation intensity for the transition and $\Gamma/2\pi$ = 50 kHz is the linewidth.

Gaussian fits to the data are shown in Fig. 4, from which the peak loading rates, linewidths and line centers are determined. The ratios of the peak loading rates for the different isotopes ($^{134}$Ba:$^{136}$Ba:$^{138}$Ba) are found to be (2:8:75), which correspond well to the natural fractional abundances of the isotopes, 2% 8%, 72%, respectively. The shifts of the line centers of the loading spectra relative to the $^{138}$Ba absorption line are found to be 103(7), 95(5) and 4(5) MHz for the $^{134}$Ba:$^{136}$Ba:$^{138}$Ba ions, respectively, not far from the known isotope shifts of 122, 109 and 0 MHz [23].

The observed linewidths of the loading spectra are 100 MHz FWHM, which is much wider than the 50 kHz natural linewidth of the 791 nm intercombination line [24], the estimated transit broadening of ~1 MHz, and the power broadened linewidth of 7.6 MHz. Although the 791 nm laser beam is aligned orthogonal to the axis of the atomic beam to minimize Doppler broadening, there is residual Doppler broadening due to the divergence of the atomic beam (δθ~0.3 rad), Using the most probable velocity of the atomic beam $v$ = 320 m/s for an oven temperature of 300 °C yields an estimate of the Doppler broadening of 102 MHz FWHM, in good agreement with the observed linewidth.

Even with this Doppler broadened linewidth, which could be reduced by a factor of 10 with a straightforward modification of the oven geometry, it is possible to achieve good isotope selectivity between the two most abundant isotopes, $^{136}$Ba and $^{138}$Ba. It is also possible to load the $^{134}$Ba isotope with good discrimination from $^{138}$Ba, however, the $^{134}$Ba-$^{136}$Ba isotope shift of only 13 MHz prevents good selectivity between these two isotopes with our current geometry.

Even in the case where the undesired isotope is loaded, it is possible to purify the ion chain after loading by exploiting the isotope shifts of the ion cooling lights. The shifts of the cooling light are small enough that if the 493 nm and 650 nm lasers are tuned to the point of optimal brightness for $^{138}$Ba, then both $^{136}$Ba and $^{134}$Ba are dimly visible in the ion chain. This is evident in Fig. 2(b). Starting with an ion chain containing ions of both $^{138}$Ba and $^{136}$Ba, the $^{138}$Ba ions may be driven out of the trap by tuning the cooling light to the point of optimal brightness for $^{136}$Ba. The 493 nm and 650 nm light is then blue detuned from transition in $^{138}$Ba and after ~20 s any $^{138}$Ba ions initially loaded have been heated out of the trap. $^{134}$Ba ions may similarly be distilled from a chain containing any $^{138}$Ba or $^{136}$Ba.

The isotope selectivity of the photoionization loading has also been investigated for the odd isotopes, $^{135}$Ba and $^{137}$Ba. These isotopes are particularly important for quantum information applications because of the hyperfine structure of the ground states due to the non-zero nuclear spin (I = 3/2) of these isotopes. Of course, the additional structure necessitates additional laser frequencies to address the different hyperfine states in both the ground and excited states. For the current study, we instead employ sympathetic cooling of the odd isotopes using laser-cooled $^{138}$Ba ions.

The technique is as follows: first, several $^{138}$Ba ions are loaded into the trap, cooled into a linear crystal and counted. Next, the photoionization loading is attempted with the odd isotopes. The odd isotope ions are cooled sympathetically into the ion chain and then their number is determined by counting how many dark spots the ion chain contains.

The results for the odd isotope loading are shown in Fig. 5. Although counting the dark spots precludes differentiating between the two odd isotopes, $^{135}$Ba and $^{137}$Ba, the two-peaked feature of the data suggests isotopically selective loading of these two isotopes as well. The data are fit to the sum of two Gaussian curves; the peaks are separated by 140 MHz and the relative heights are in agreement with the natural abundance ratio of 1:1.7. For these data, the power of the 791 nm photoionization laser was to 0.5 mW.

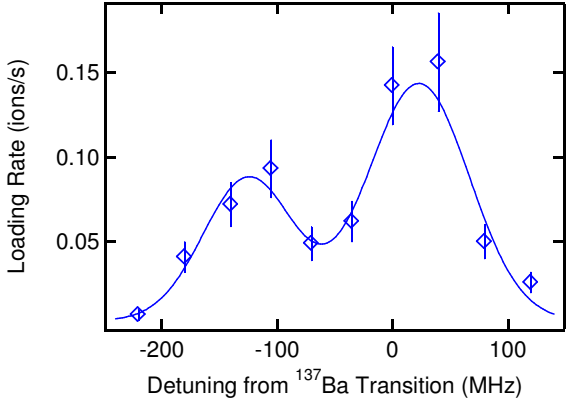

FIG. 5 (Color Online) Loading rates for the odd isotopes, $^{135}$Ba and $^{137}$Ba vs. detuning of the 791 nm photoionization laser relative to the $^{137}$Ba absorption line. The solid curve is a fit to a sum of two Gaussian curves.

We have also investigated the loading rate dependence on the intensity of the 791 nm laser beam, the results of which are shown in Fig. 6. The increased loading for increasing power is attributed to power broadening of the transition, which scales as $\sqrt{P}$, where $P$ is the power. Since the natural linewidth of the intercombination transition is small compared with the 102 MHz Doppler broadened width, power broadening will excite an increasing fraction of the velocity classes in the atomic beam. This fraction will increase linearly with respect to the power broadened linewidth up to the limit of the Doppler broadening. Indeed, a $\sqrt{P}$ curve fits the data fits well.

Within this simple model, the loading rate will saturate when the power broadened linewidth is sufficient to excite all of the velocity classes in the atomic beam. In the current set-up, this would require 17 mW and would yield a maximum loading rate of 10 ions/s. This provides an estimate for the potential loading rate with a true Doppler-free geometry for the present laser intensities.

Finally, we discuss a particularly simple, low cost and efficient technique for loading ion traps that, to our knowledge, has not been reported previously. It is observed that ions can be loaded into the trap by simply shining a halogen lamp onto the ion trap. The measured loading rate is 2.4 ions/s, which is comparable to the photoionization loading rate. The lamp employed (Thorlabs UV75) emits light over the wavelength range of 200-540 nm and is typically used for curing UV epoxy. Since the vacuum windows do not transmit light below 300 nm, the light responsible for the loading is in the 300 to 540 nm range. We have determined that the underlying mechanism behind the loading is the photoelectric effect. Indeed, we measure a photoelectric current that is collected by a nearby positively biased conductor. The measured photocurrent is comparable to the current created by an auxiliary hot-wire based electron beam used for earlier studies.

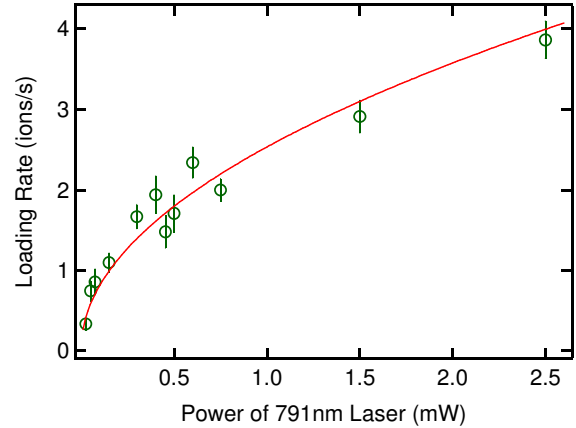

FIG. 6 (Color Online): Loading rate as a function of 791 nm laser power. The data closely follows a line proportional to the square root of the 791 nm laser power.

Nominally, the work function of the stainless steel electrodes is too high (~5 eV) to eject electrons with these wavelengths. However, the work function of bulk barium is much lower (2.7 eV) [25], and it is likely that the photoelectrons come from barium deposited on the trap surfaces over time. A small loading rate (0.005 ions/s) was found to be due to the illumination of the $N_2$ laser alone. Although it is not possible to measure a photo-electron current in this case due electrical noise generated by the laser, we believe that the underlying loading mechanism is the same.

TABLE I. Ion loading rates for the three techniques investigated. All rates are for the oven at 300 ºC.

| Technique | Rate (ions/s) |
| --- | --- |
| $N_2$ laser alone | 0.005(3) |
| e- beam | 0.014(6) |
| UV Lamp | 2.4(3) |
| PI (0.75 mW, $^{138}$Ba) | 2.0(1) |
| PI (max, theory) | 9.9(1) |

An important figure of merit for these studies is the relative efficiencies of the different loading techniques. For an oven temperature of 300 ºC, the relative efficiencies of e- beam bombardment, photoelectric loading and photoionization loading are 1:160:185$R$, respectively, where $R$ is the natural abundance (0 to 1) of the isotope of interest. The comparison is for a 791 nm laser power of 0.75 mW. It is noted that the UV lamp is 100× more efficient at loading the trap than the hot wire e-beam, and furthermore the ions loaded using the UV lamp tend to cool down much more quickly. The loading rates are summarized in Table 1.

In conclusion, we have demonstrated a simple scheme for photoionization loading of barium into an ion trap. Loading rates exceeding 4 ions/s are achieved for the most abundant isotope. By varying the detuning of the 791 nm laser, we achieved isotope selective loading. Using this technique, we load all available stable isotopes with abundances greater than 1%, including the even isotopes, $^{134}$Ba, $^{136}$Ba, $^{138}$Ba, and the odd-isotopes $^{135}$Ba, $^{137}$Ba. Additionally, we demonstrated a particularly simple and inexpensive loading technique utilizing photoelectrons generated by a UV curing lamp.

This work was supported by the NSF Grant PHY-0326315. We would like to thank Boris Blinov and Christopher Monroe for help with the early stages of this work, and Jeff Sherman for valuable discussions. Also, we would like to thank David Ames at SRS for the loan of a N$_2$ gas laser.